# Efficient Volumetric Method of Moments for Modeling Plasmonic Thin-Film Solar Cells with Periodic Structures

Zi He[1], Ji Hong Gu[1], Wei E. I. Sha[2], and Ru Shan Chen[1]

*1. Department of Communication Engineering, Nanjing University of Science and Technology, Nanjing, 210094, China*

*2. Key Laboratory of Micro-Nano Electronic Devices and Smart Systems of Zhejiang Province, College of Information Science and Electronic Engineering, Zhejiang University, Hangzhou 310027, China.*

*\*eerschen@njust.edu.cn*

**Abstract:** Metallic nanoparticles (NPs) support localized surface plasmon resonances (LSPRs), which enable to concentrate sunlight at the active layer of solar cells. However, full-wave modeling of the plasmonic solar cells faces great challenges in terms of huge computational workload and bad matrix condition. It is tremendously difficult to accurately and efficiently simulate near-field multiple scattering effects from plasmonic NPs embedded into solar cells. In this work, a preconditioned volume integral equation (VIE) is proposed to model plasmonic organic solar cells (OSCs). The diagonal block preconditioner is applied to different material domains of the device structure. As a result, better convergence and higher computing efficiency are achieved. Moreover, the calculation is further accelerated by two-dimensional periodic Green's functions. Using the proposed method, the dependences of optical absorption on the wavelengths and incident angles are investigated. Angular responses of the plasmonic OSCs show the super-Lambertian absorption on the plasmon resonance but near-Lambertian absorption off the plasmon resonance. The volumetric method of moments and explored physical understanding are of great help to investigate the optical responses of OSCs.

## 1. Introduction

In the last decade, the study of nanostructures has been paid much attention [1-3]. Thin-film organic solar cells (OSCs) have been widely used since they are potentially cost-effective in photovoltaic applications. In view of short exciton diffusion length and low carrier mobility, many methods have been applied to enhance the weak optical absorption of OSCs due to the thin active layer designs [4-6]. One of the most popular ways is to incoporate the metallic nanostructures, which excite surface plasmon modes and trap the light at the absorbing layer. Apart from using metallic nanoparticles (NPs), adopting periodic structures is also an efficient approach to enhance absorption efficiency [7-8]. However, the optimized design of OSCs suffers from a tremendous challenge caused by the huge computational resources needed. Two-dimensional semi-analytical methods have been employed to design the shapes and distributions of the plasmonic NPs in OSCs [9-11]. In [9], the near-field multiple scattering effects at both the carrier transport and active layers were investigated, respectively. Then the light absorption was computed by using the two-dimensional finite-element method (FEM) [10]. More recently, plasmonic hybridization from Ag nanomaterials of different shapes has been modelled to achieve broadband absorption [11]. However, few numerical works reported full-wave simulations and analyses of OSCs embedded with the plasmonic NPs. Therefore, the study on the efficient numerical computation of plasmonic OSCs has great significances for high-performance photovoltaics.

Three-dimensional full-wave numerical methods have been widely adopted for the electromagnetic analyses and designs. In recent years, several full-wave numerical techniques have been successfully applied to model OSCs, such as the finite-difference frequency-domain (FDFD) [12-13], finite-element boundary-integral (FE-BI) [16], and finite-difference time-domain (FDTD) methods [14-15]. For the FDFD and FDTD approaches, the staircase errors will be introduced since the structures are discretized with rectangular boxes. Moreover, a large number of summation terms in the Lorentz-Drude model are required to describe the dispersive optical materials for the time-domain numerical techniques. In this way, the computational resources are increased significantly. Furthermore, for the FE-BI method, the impedance matrix is ill-conditioned and thus the convergence is slow. In addition, the periodic boundary conditions are not easy to handle when the incident light is oblique for the FDTD method [17-18]. Hence, the frequency-domain integral equation method exhibits important advantages over the above-mentioned numerical methods for modeling OSCs with periodic structures.

In this work, Maxwell's equations are rigorously solved by using the volume integral equation (VIE). Due to the surface-wave nature of surface plasmon polaritons, the convergence of iterative methods becomes very slow. Thus, the diagonal block preconditioner is introduced to each domain of solar cell structures to obtain faster convergence. Additionally, the two-dimensional periodic Green's functions are utilized, which is equivalent to the Bloch boundary conditions implemented in the FEM or FD method. Thus, only a unit cell needs to be modeled instead of whole space of the OSC device. The incident plane waves covering the wavelength range of 400-800 nm are assumed as the sunlight illumination. We theoretically study the light absorption enhancement for versatile periodic structures and incident angles. Particularly, angular responses of the plasmonic OSCs are investigated in detail. The work is important for designing and optimizing high-efficiency OSCs.

The reminder of the paper is organized as follows: Section 2 presents the theory of the VIE, two-dimensional periodic Green's function and domain decomposition-based diagonal block preconditioner. Section 3 shows numerical results that demonstrate the efficiency of the proposed method and studies the dependences of light absorption on the wavelengths and incident angles. Finally, a conclusion is given in Section 4.

## 2. Theory and formulations

In this section, the VIE is formulated in terms of the polarization current firstly. Then the two-dimensional periodic Green's functions are introduced to accelerate the calculation. Finally, a domain decomposition-based diagonal block preconditioner is proposed to improve the convergence for multi-layered and multi-domain device structures.

*2.1 Theory of VIE*

Assume that $\exp(j\omega t)$ is the time dependence of the fields and all the variables are defined in the Cartesian coordinates. It should be noted that the non-magnetic material is considered in this paper. Then the volume integral equation [26] can be established as:

$$\frac{\mathbf{J}_v(\mathbf{r})}{j\omega\varepsilon_0(\varepsilon_r - 1)} + j\omega\mathbf{A}(\mathbf{r}) + \nabla\Phi(\mathbf{r}) =$$
$$\frac{\mathbf{J}_V(\mathbf{r})}{j\omega\varepsilon_0(\varepsilon_r - 1)} + j\omega\mu_0 \int_V g(\mathbf{r},\mathbf{r}')\mathbf{J}_v(\mathbf{r}')dV' \quad (1)$$
$$- \frac{1}{j\omega\varepsilon_0}\nabla\int_V g(\mathbf{r},\mathbf{r}')\nabla'\cdot\mathbf{J}_v(\mathbf{r}')dV' = \mathbf{E}^i(\mathbf{r})$$

where $g(\mathbf{r},\mathbf{r}') = \frac{e^{-jk|\mathbf{r}-\mathbf{r}'|}}{4\pi|\mathbf{r}-\mathbf{r}'|}$ is the scalar Green function, $\mathbf{J}_v$ is the volumetric polarization current, and the permittivity and permeability in free space are $\varepsilon_0$ and $\mu_0$, respectively. The position-dependent relative permittivity of the inhomogeneous solar cell is $\varepsilon_r(\mathbf{r})$, and $\mathbf{E}^i(\mathbf{r})$ is the incident field.

The Schaubert–Wilton–Glisson (SWG) basis functions are employed to expand the volumetric polarization currents. Then the matrix equation can be obtained by using the Galerkin's test method [22].

$$
\begin{aligned}
&(j\omega)^2 \mu_0 \int_{v_m} dv_m \mathbf{f}_m^v(\mathbf{r}) \int_{v_n'} \kappa_n(\mathbf{r}') \mathbf{f}_n^v(\mathbf{r}') \cdot g(\mathbf{r},\mathbf{r}') dv_n' \\
&+ \frac{1}{\varepsilon_0} \int_{v_m} dv_m \nabla \cdot \mathbf{f}_m^v(\mathbf{r}) \int_{v_n'} \kappa_n(\mathbf{r}') \nabla' \cdot \mathbf{f}_n^v(\mathbf{r}') \cdot g(\mathbf{r},\mathbf{r}') dv_n' \\
&- \frac{1}{\varepsilon_0} \int_{v_m} dv_m \nabla \cdot \mathbf{f}_m^v(\mathbf{r}) \int_{v_n'} \left( \mathbf{f}_n^v(\mathbf{r}') \nabla' \kappa_n(\mathbf{r}') \right) \cdot g(\mathbf{r},\mathbf{r}') dv_n' \\
&- \frac{1}{\varepsilon_0} \int_{\Omega_m} \mathbf{f}_m^v(\mathbf{r}) \cdot \mathbf{n} d\Omega_m \int_{v_n'} \kappa_n(\mathbf{r}') \nabla' \cdot \mathbf{f}_n^v(\mathbf{r}') \cdot g(\mathbf{r},\mathbf{r}') dv_n' \\
&+ \frac{1}{\varepsilon_0} \int_{\Omega_m} \mathbf{f}_m^v(\mathbf{r}) \cdot \mathbf{n} d\Omega_m \int_{v_n'} \left( \mathbf{f}_n^v(\mathbf{r}') \nabla' \kappa_n(\mathbf{r}') \right) \cdot g(\mathbf{r},\mathbf{r}') dv_n' \\
&+ \int_{v_m} dv_m \mathbf{f}_m^v(\mathbf{r}) \frac{1}{\varepsilon_n(\mathbf{r}')} \mathbf{f}_m^v(\mathbf{r}') = \int_{v_m} \mathbf{E}^i(\mathbf{r}) \cdot \mathbf{f}_m^v(\mathbf{r}) dv_m
\end{aligned}
\quad (2)
$$

where $\kappa_n(\mathbf{r}') = \dfrac{\varepsilon_0 \varepsilon_r(\mathbf{r}') - \varepsilon_0}{\varepsilon_0 \varepsilon_r(\mathbf{r}')}$, $\mathbf{f}^v$ is the SWG basis function, and $\Omega$ stands for the outside interface of the tetrahedron $v$.

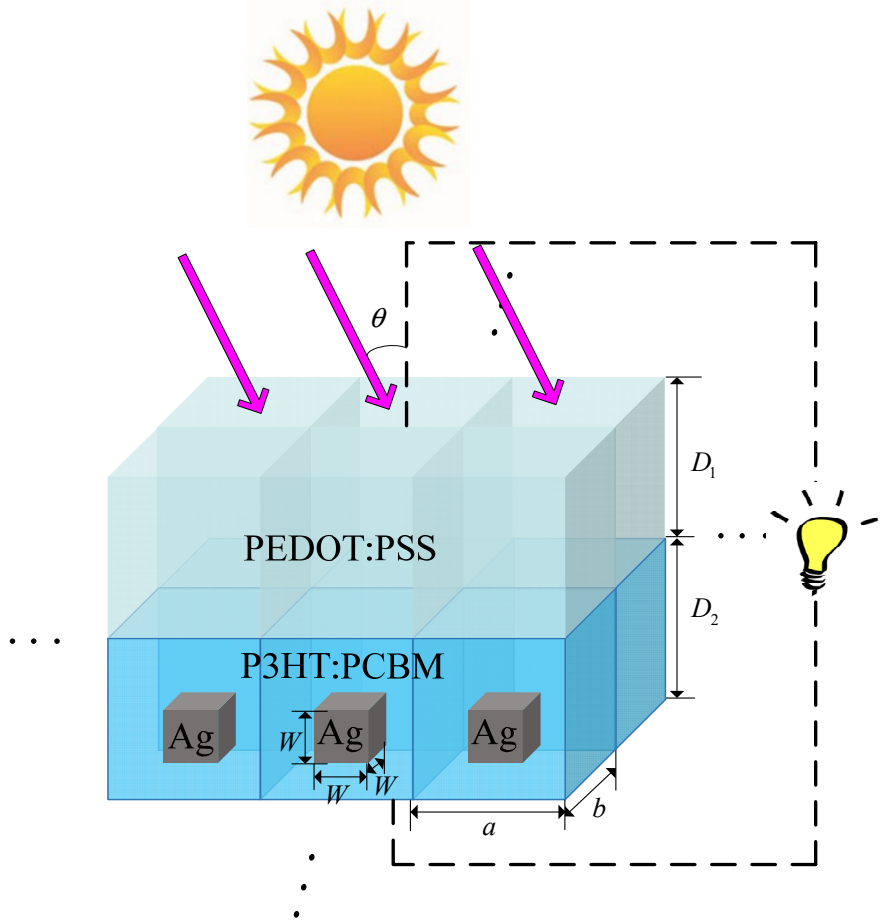

Fig. 1. The schematic of the plasmonic OSC. Periodic silver (Ag) NP array is embedded at the active layer (P3HT:PCBM) of the OSCs. The PEDOT:PSS layer is a spacer layer. The device is illuminated by sunlight with an incident angle of $\theta$. The geometry of the device structure is: The cell size is $a \times b \times D_1$ for the PEDOT:PSS domain, $a \times b \times D_2$ for the P3HT:PCBM domain and $W \times W \times W$ for the Ag domain.

*2.2 Two-Dimensional Periodic Green's Functions*

Fig. 1 shows the schematic of the plasmonic OSC, which is periodic along the *x* and *y* directions. The lattice constants along the *x* and *y* axes are *a* and *b*, respectively. The incident angle of the sunlight is set to be $\theta$. Then the two-dimensional periodic Green's functions (PGF) can be derived as [19-20]

$$g_p = \sum \frac{e^{-jk_0 R_{mn}}}{4\pi R_{mn}} e^{j(k_x^i ma + k_y^i nb)} \tag{3}$$

where $k_x^i = k_0 \sin\theta \cos\varphi$ and $k_y^i = k_0 \sin\theta \sin\varphi$. According to the Bloch-Floquet theorem, $k_x^i$ and $k_y^i$ are the phase shifts between adjacent cells along the *x* and *y* directions, and $(\theta, \varphi)$ is the incident angle. $R_{mn}$ is the distance between the source and observation points, i.e.

$$R_{mn} = \sqrt{(x - x' + ma)^2 + (y - y' + nb)^2 + (z - z')^2} \tag{4}$$

It should be noted that the calculation of the periodic Green's functions can be accelerated by the Ewald's method [21, 27-28]. The Ewald's method splits the periodic Green's functions into a sum of the spatial and spectral series. More specifically, the spatial series can be calculated with the Ewald identity [21] and the spectral series can be obtained by using the Poisson summation formula [25]. In this way, both the spatial and spectral series achieve good convergence.

*2.3. Domain Decomposition-based Diagonal Block Preconditioner*

As shown in Fig. 1, without loss of generality, there are three types of materials for a unit cell of the plasmonic OSC, including the spacer layer (PEDOT:PSS) domain, active layer (P3HT:PCBM) domain and plasmonic NP (Ag) domain. The cell size is $a \times b \times D_1$ for the PEDOT:PSS domain, $a \times b \times D_2$ for the P3HT:PCBM domain and $W \times W \times W$ for the Ag domain. Then the impedance matrix can be split into small submatrices with respect to different material domains. Thus, the matrix equation can be rewritten as

$$\begin{bmatrix} A_{11} & A_{12} & A_{13} \\ A_{21} & A_{22} & A_{23} \\ A_{31} & A_{32} & A_{33} \end{bmatrix} \begin{bmatrix} x_1 \\ x_2 \\ x_3 \end{bmatrix} = \begin{bmatrix} b_1 \\ b_2 \\ b_3 \end{bmatrix} \tag{5}$$

where $A_{ii}$ $(i = 1, 2, 3)$ describes the self-interaction of the *i*-th domain, and $A_{ij}$ $(i = 1, 2, 3, j = 1, 2, 3)$ represents the mutual interaction between the *i*-th and *j*-th domains.

Considering the convergence will become worse for different dispersive media, we use the block diagonal preconditioner to handle each dispersive medium region. Moreover, the block diagonal preconditioner is easy to be implemented and has a good performance. Then, the diagonal block preconditioner is respectively constructed for $A_{11}, A_{22}, A_{33}$ as follows:

$$\begin{bmatrix} A_{11}^{-1} & & \\ & A_{22}^{-1} & \\ & & A_{33}^{-1} \end{bmatrix} \begin{bmatrix} A_{11} & A_{12} & A_{13} \\ A_{21} & A_{22} & A_{23} \\ A_{31} & A_{32} & A_{33} \end{bmatrix} \begin{bmatrix} x_1 \\ x_2 \\ x_3 \end{bmatrix} = \begin{bmatrix} A_{11}^{-1} b_1 \\ A_{22}^{-1} b_2 \\ A_{33}^{-1} b_3 \end{bmatrix} \quad (6)$$

Compared to Eq. (5), the convergence of the preconditioned Eq. (6) is significantly improved. Consequently, the three-dimensional full-wave analyses of the OSCs incorporating metallic NPs can be numerically implemented with a high efficiency.

## 3. Numerical results

All the numerical results are produced by the computer of Intel Xeon E7-4850 CPU equipped with 8GB RAM. The incident plane waves with the wavelengths ranging from 400 nm to 800 nm are assumed as the sunlight illumination. The light absorption is defined as

$$A(\lambda) = \omega \int_v \text{Im}(\varepsilon(\lambda, \mathbf{r})) |\mathbf{E}(\lambda, \mathbf{r})|^2 \, dv \quad (7)$$

where $\lambda$ is the wavelength, $\text{Im}(\varepsilon)$ denotes the imaginary part of the complex permittivity for the active organic materials, and $\mathbf{E}$ represents the electric field at the active layer. It should be noted that the integral is only made at the active material domain of the OSC (excluding the metallic domain). Then, absorption enhancement can be obtained by dividing the light absorption of the OSC with the NPs by that of the OSC without the NPs.

In the beginning, the convergence performance of the domain decomposition-based diagonal block preconditioner is tested under the wavelength of 400 nm condition. The device structure of the OSC is depicted in Fig. 1 and only six unit cells are shown. Ag NPs are embedded into the active layer. *W* and *a (a=b)* are chosen as 20 nm and 40 nm, respectively. And the thickness is 20 nm for the spacer layer, 60 nm for the active layer. As illustrated in Fig. 2, regarding the two-layer OSC, the convergence curves of the generalized minimal residual (GMRES) iteration method [23] with and without the preconditioner are given. Moreover, Fig. 3 depicts the relative errors of the current coefficients, which can be defined as

$$\varepsilon_{re} = \frac{\|J_{Pre} - J_{No-pre}\|_2}{\|J_{No-pre}\|_2} \quad (8)$$

where $J_{Pre}$ and $J_{No-pre}$ are the current coefficients obtained with and without the preconditioner, respectively.

As shown in Fig. 3, the relative errors of all the unknowns are less than 12%. It can be concluded that the computational efficiency can be improved with encouraging accuracy by using the proposed preconditioner technique.

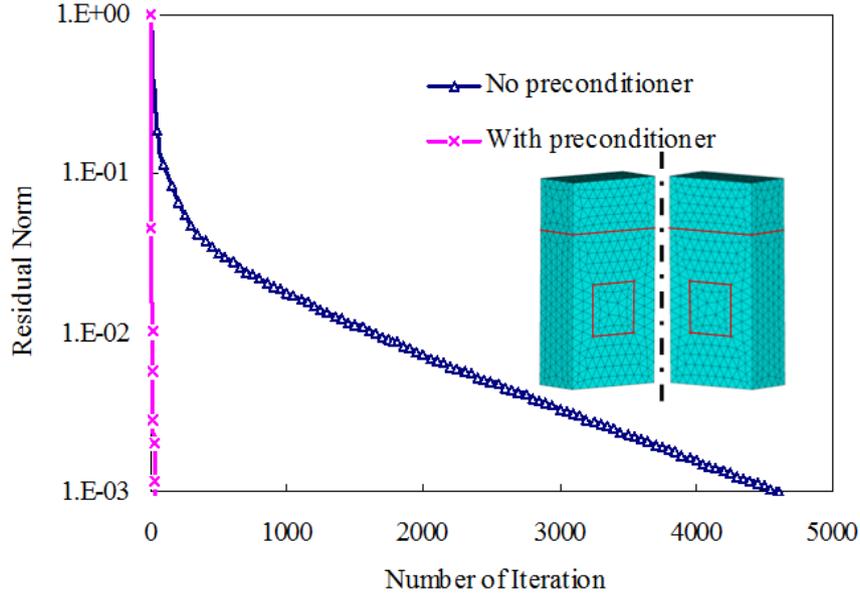

Fig. 2. Residual convergence curves of the GMRES iteration for the plasmonic OSC. The cross and triangle lines are the convergence results for the VIE methods with and without the preconditioners, respectively.

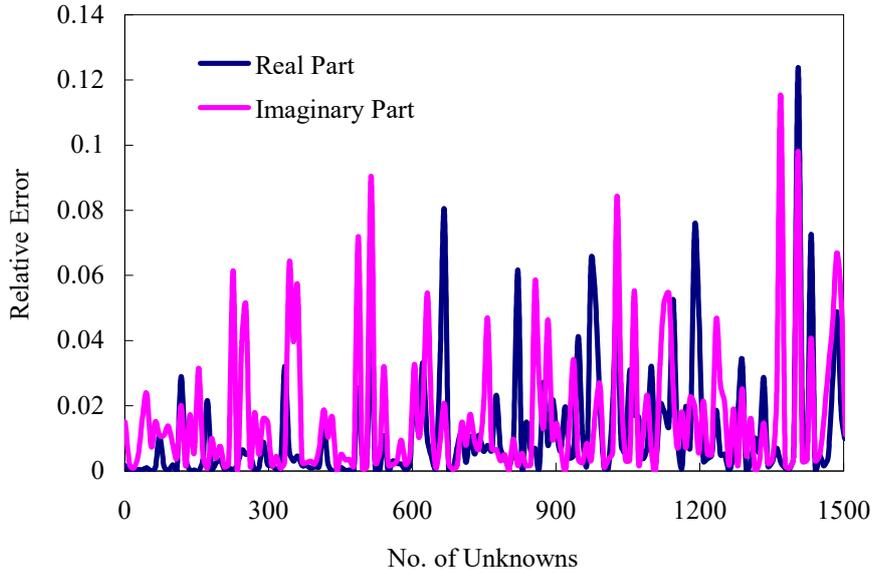

Fig. 3. Relative errors of the complex current coefficients by the VIE methods with and without the preconditioners.

Then we optimize the lattice constant $a$ $(a=b)$ to maximize the absorption enhancement with $W$=40 nm at the perpendicular incidence. The thickness of the active layer is 60 nm and that of the spacer layer is 20 nm. As shown in Fig. 4, the absorption spectra of the OSC are demonstrated as a function of the lattice constant. As the lattice constant decreases, the plasmonic coupling and hybridization between adjacent Ag NPs become stronger. The largest absorption enhancement factor up to 32 can be obtained for $a$ =60 nm. However, a very small lattice constant (a =50 nm) reduces the volume of active materials (P3HT:PCBM) significantly; and thus lowers the optical absorption. The absorption spectra show three peaks around 500 nm, 600 nm, and 750 nm as depicted in Fig. 4. The waveguide mode (at the peak 1, E-field is bound between the top and bottom interfaces of the P3HT:PCBM layer), blue-shifted transverse plasmonic mode (at the peak 2), and red-shifted longitudinal plasmonic mode (at the peak 3) are excited from short wavelengths to long wavelengths, respectively. The strongest enhancement is gained by the longitudinal plasmonic mode with the resonating wavelength of 750 nm. To further understand the optical enhancement mechanisms, the wavelength-dependent absorption profiles at the active layer are illustrated in Fig. 5 with $a$ =60 nm. Although the Ag NPs show large metallic loss, the plasmonic near fields still penetrate into the active layer and absorbed by organic active materials.

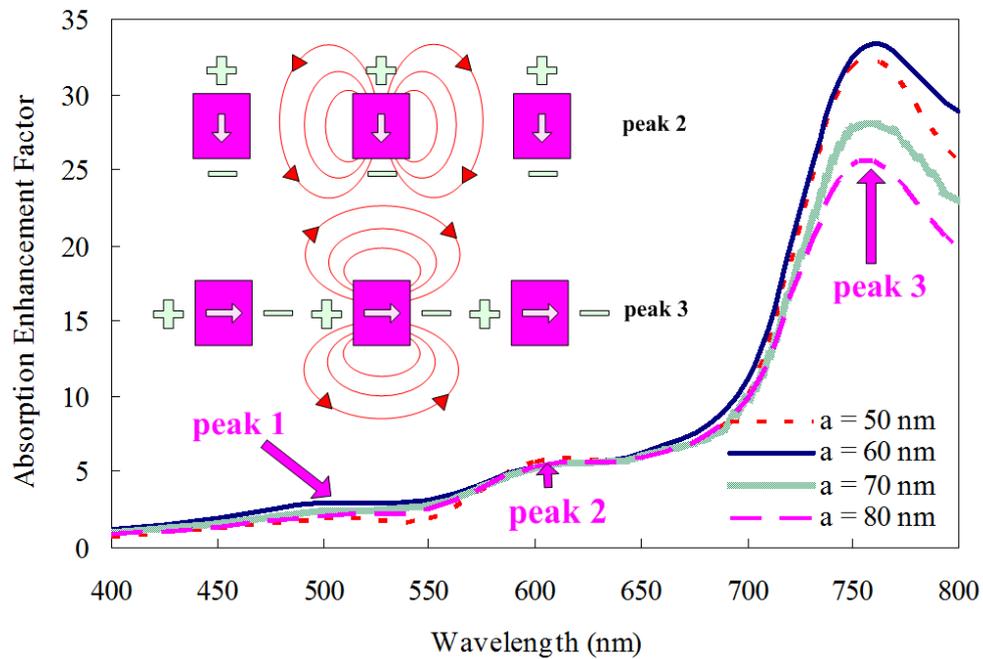

Fig. 4. Absorption spectra of the plasmonic solar cell as a function of the lattice constant. The absorption is calculated only from the P3HT:PCBM active material. The arrows denote the resonance peaks. The peaks 1, 2, 3 are related to the excited waveguide mode, transverse plasmonic mode, and longitudinal plasmonic mode. The coupling mechanisms of the transverse mode (peak 2) and longitudinal mode (peak 3) are shown in the inset.

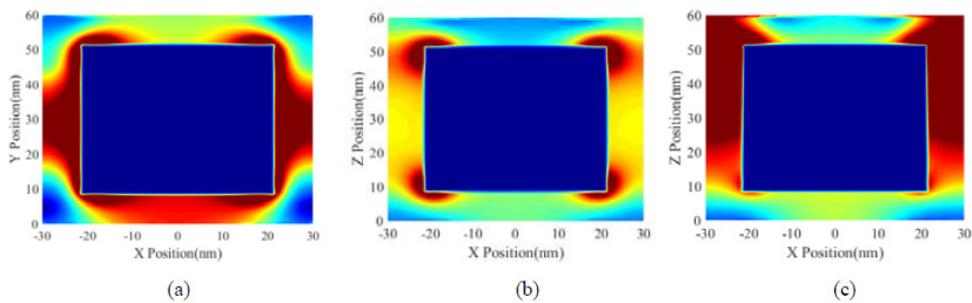

Fig. 5. Absorption profiles of active layer at the three resonance wavelengths of Fig. 4. The lattice constant is a=60 nm. (a) $\lambda$ = 500 nm (peak 1); (b) $\lambda$ = 600 nm (peak 2); (c) $\lambda$ = 750 nm (peak 3).

Finally, the performance of the OSC ($W$=40 nm, $a$=$b$=60 nm, $D_1$=20 nm, $D_2$=60 nm) is theoretically modelled over a wide range of incident angles, which is essential to the practical solar cell devices [24]. Fig. 6 shows the wavelength-dependent angular response of the plasmonic OSC. The absorption enhancement factors decrease as the incident angle increases. On the longitudinal plasmon resonance wavelength (750 nm), the normalized absorption, as the incident angle increases, shows the super-Lambertian feature, i.e. it decays much slower than the ideal Lambertian absorption curve of $\cos(\theta)$. On the transverse plasmon resonance wavelength (600 nm), the absorption decays faster but still shows the super-Lambertian absorption. In contrast, on the waveguide resonance wavelength, which is off the plasmon resonance wavelengths, the angular response is more close to the ideal Lambertian curve. Thus, the plasmonic resonances, compared to the waveguide resonance, are more insensitive to the incident angle.

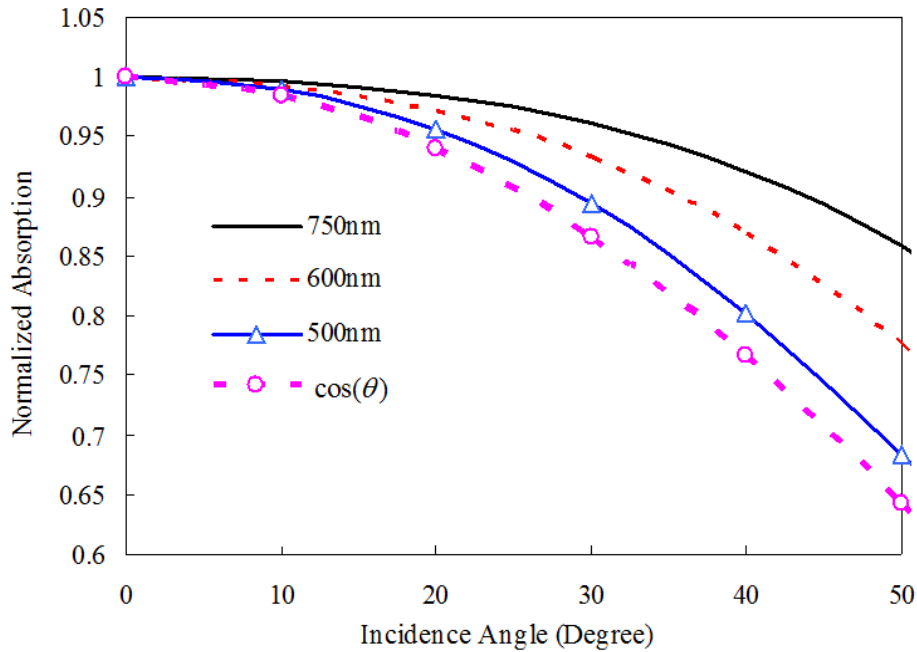

Fig. 6. Wavelength-dependent angular responses of the plasmonic solar cell. The ideal Lambertian curve is given for comparisons. The 500 nm, 600 nm, and 750 nm correspond to the three resonance wavelengths of Fig. 4.

## 4. Conclusion

A volumetric method of moments is proposed as a full-wave solution to model thin-film solar cells incorporating periodic plasmonic structures. By introducing the two-dimensional periodic Green's functions, the calculation is only needed to be implemented in a unit cell. Therefore, the computational resources can be saved significantly. Particularly, a domain decomposition-based diagonal block preconditioner is applied at different material domains to overcome the bad convergence of impedance matrix equation caused by plasmonic effects.

Numerical examples are presented to demonstrate the efficiency of the proposed method. Most importantly, we found that the angular responses of plasmonic solar cells significantly break the Lambertian absorption limits on the plasmon resonance wavelengths but show near-Lambertian absorption off the plasmon resonance wavelengths. Our work offers a promising mathematical approach and deep physical understanding to the optical designs of nanostructured OSCs.

**Funding**

This work was supported in part by Natural Science Foundation of 61701232, Jiangsu Province Natural Science Foundation of BKs20170854, the Young Elite Scientists Sponsorship Program by CAST of 2017QNRC001, the China Postdoctoral Science Foundation of 2017M620861 and 2018T110127, the Fundamental Research Funds for the central Universities of No.30917011317, and the State Key Laboratory of Millimeter Waves of K201805. The project was also supported by Thousand Talents Program for Distinguished Young Scholars of China (588020-X01801/009).

**Acknowledgment**

We thank Prof. Weng Cho Chew from Purdue University for helpful discussions about theoretical details.